\documentclass{emulateapj}
\usepackage{apjfonts}
\newcommand{\etal}{{et al}\/.}

\begin{document}
\slugcomment{accepted for publication in ApJ Letters}
\shorttitle{Gemini imaging of a z=10 candidate}
\shortauthors{M. N. Bremer \etal}
\title{Gemini H-band imaging of the field of a z=10 candidate}
\author{M.N. Bremer}
\affil{Department of Physics, University of Bristol, Tyndall Avenue,
Bristol BS8 1TL, UK}
\author{Joseph B. Jensen}
\affil{Gemini Observatory, 950 N. Cherry Ave., Tucson, AZ 85719}
\author{M. D. Lehnert, N.M. F\"{o}rster Schreiber}
\affil{Max-Planck-Institut f\"ur extraterrestrische Physik,
Giessenbachstra\ss e, 85748 Garching bei M\"{u}nchen, Germany}
\and
\author{Laura Douglas}
\affil{Department of Physics, University of Bristol, Tyndall Avenue,
Bristol BS8 1TL, UK}

\begin{abstract}

We present a deep $H$-band image of the field of a candidate $z{=}10$
galaxy magnified by the foreground ($z{=}0.25$) cluster Abell 1835. The
image was obtained with NIRI on Gemini North to better constrain the
photometry and investigate the morphology of the source. The image is
approximately one magnitude deeper and has better spatial resolution
(seeing was  0.4-0.5 arcsec) than the existing $H$-band image obtained
with ISAAC on the VLT by \cite{pello04}.  The object is not detected
in our new data. Given the published photometry ($H_{AB}$=25.0), we
would have expected it to have been detected at more than $\sim7\sigma$
in a 1.4 arcsec diameter aperture.  We obtain a limit of $H_{AB}$$>$26.0
(3$\sigma$) for the object. A major part of the evidence that this object
is at $z{=}10$ was the presence of a strong continuum break between the $J$
and $H$ band, attributed to absorption of all continuum shortward
of 1216 \AA~ in the rest-frame of the object. Our $H-$band non-detection
substantially reduces the magnitude of any break and therefore weakens the
case that this object is at $z{=}10$.  Without a clear continuum break,
the identification of an emission line at 1.33745$\mu$m as Ly$\alpha$
at $z\approx10$ is less likely.  We show that the width and flux of
this line are consistent with an alternative emission line such as
[OIII]$\lambda$5007 from an intermediate redshift HII/dwarf galaxy.

\end{abstract} \keywords{cosmology: observations - early universe -
galaxies: distances and redshifts - galaxies: evolution - galaxies:
formation}

\maketitle
\section{Introduction}
\label{intro}
Eight-meter class telescopes such as Gemini, VLT and Keck have opened
up the Universe at $z>5$ for detailed study. The first $z>5$ galaxy
was discovered in 1998 \citep{Dey98}. In the past two years the number
of confirmed and candidate $z>5$ galaxies has grown substantially,
\citep[see {\it e.g.,}][]{L03,bremer04, S04a, S04b, bunker03, ajiki03,
rhoads03, bouwens04, hu04}. These have been discovered using a variety
of techniques including slitless spectroscopy, narrow-band imaging
surveys for Ly$\alpha$ emitters, and broad-band photometry followed by
spectroscopy to identify UV-bright Lyman dropouts.

Until recently, most searches for distant galaxies have concentrated
on redshifts out to $z\sim6.6$. Beyond 7000 \AA~ the sky emission is
increasingly dominated by OH emission bands, making spectroscopic
identification of Ly$\alpha$ emission more and more difficult, except
in the gaps between the OH bands.  As a consequence the redshift
distribution of $z>5$ galaxies with spectroscopic redshifts is
non-uniform, mostly reflecting the wavelength distribution of these
gaps. Nevertheless, with 8m telescopes it is entirely possible to
unambiguously identify galaxies at $z\sim5-6$, \cite[see e.g., the
spectra in][]{L03, ando04}. Beyond 9000 \AA~ the sky shows a higher
density of bright telluric lines and a brighter continuum. As the
typical $z\sim5-6$ galaxy has a continuum magnitude of AB$>25$ and a
roughly flat intrinsic f$_{\nu}$ spectrum (zero color in AB), this
makes identifying candidates at even higher redshifts increasingly
difficult.

Nevertheless, surveys for even more distant galaxies are
particularly important for our understanding of reionization. The
likely presence of Gunn-Peterson troughs in the spectra of $z>6$
SDSS quasars \citep{becker01, djorgovski01, fan02} indicate that the
reionization of hydrogen in the IGM ended at $z\ga6$. Reionization may
have begun much earlier; analysis of Wilkinson Microwave Anisotropy
Probe (WMAP) first-year temperature and polarization data indicates
possible substantial reionization at $z>10$ \citep{kogut03}. As the
UV emission from early star-forming galaxies is thought to be a major
(and possibly dominant) source of ionizing photons, the evolution in the
number and luminosity of such galaxies is likely to be directly linked
to the reionization history of the IGM. Based on work carried out thus
far to $z\sim 6$ it appears that the Universal UV luminosity and star
formation density declines with increasing redshift beyond $z\sim 3$
\citep[e.g.,][]{L03, bunker04}. If this trend continues beyond $z=6$,
it is difficult to envisage star formation in moderate mass galaxies
being responsible for ionizing the Universe at $z>6$ \citep[see
e.g.,][]{ricotti04}.

One technique developed to overcome the difficulty of detecting even more
distant galaxies is the use of gravitational lensing by an intervening
galaxy cluster  to boost their apparent brightness. This boost can be
as much as a factor of 10-100 if the galaxy happens to lie on a critical
line. \cite{santos04} proved the utility of this technique out to $z=5.6$
and \cite{kneib04} discovered a probable lensed $z\sim 7$ Lyman break galaxy
behind Abell 2218.

Recently, \cite{pello04} identified a probable highly magnified galaxy
at $z{=}10$ lying on a critical line of the cluster Abell 1835. Using
broad-band optical imaging from HST and CFHT along with near-IR imaging
and spectroscopy from ISAAC/VLT, they presented evidence for a redshift
of 10. The object was not detected in $V,R,I$ optical bands and was
formally only detected at more than 4$\sigma$ in $H$ ($H_{AB}=25.00\pm
0.25$) and about 3$\sigma$ in $K$ ($K_{s,AB}=25.51\pm 0.36$). The $J-$band
detection quoted by \citeauthor{pello04}was $J_{AB}=26.8\pm 1$. $J-$band
spectroscopy covered the wavelength range 1.162 to 1.399 $\mu$m and showed
an emission line at 1.33745 $\mu$m, detected in two separate central
wavelength settings of the spectrograph, with a combined significance
level of $4-5\sigma$.

This photometry and spectroscopy led \citeauthor{pello04}to argue that
the emission line is most likely Ly$\alpha$ at $z{=}10.0175$.  A key
part of this case was the shape of the spectral energy distribution
measured from the imaging data. The object was undetected in the optical,
showed a large break between the $J$ and $H$ bands, and had a blue
$H-K_s$ color ($H_{AB}-K_{s,AB}<0$). This spectral energy distribution
is consistent with a young stellar population observed at $z\sim10$,
with light shortward of 1216\AA~ in the rest-frame heavily absorbed by
intervening neutral hydrogen. This is consistent with the detected line
being Ly$\alpha$ at $z{=}10.0175$.

However, given the low signal-to-noise detection, the extremely
low flux of the emission line and the limited wavelength covered by
the spectroscopy, other emission line identifications are plausible
(e.g., [OIII]$\lambda$5007, one component of the [OII]$\lambda$3727
doublet, H$\alpha$).  Although \citeauthor{pello04} claim the source
is on a $z=10$ critical line for Abell 1835, which could be used as
a compelling evidence of its extreme redshift, there is no published
detailed analysis of the accuracy of the mass model. The critical lines
for other redshifts are close to those at $z=10$, so uncertainty in the
mass model can lead to the source falling on or near a critical line
of a different redshift. This uncertainty is reflected in the range of
magnifications given in \cite{pello04}.  With all of the uncertainties
in the other pieces of evidence that the source is at z=10,
the strength of the $H-$band detection is key to this interpretation;
without a strong break between $J$ and $H$, there is no other compelling
evidence that the line should be identified as Ly$\alpha$.

Therefore, given the crucial role of the $H-$band imaging in assigning a
redshift to this object, we here report on deep $H-$band images of the
field of the $z{=}10$ candidate obtained with the Near-Infrared Imager
(NIRI) on the Frederick C. Gillett Gemini Telescope (Gemini-North).
These images were obtained both to better constrain the $H-$band photometry
(the original detection was not highly significant) and to investigate
the morphology of the source with images taken under the excellent seeing
conditions that are often attainable at Gemini-North.

\section{Data and analysis}

\subsection{Observations and Data Reduction}

NIRI observations of Abell 1835 were carried out using Director's
Discretionary Time on the nights of 30 May and 6 Jun 2004 UT.  NIRI is
an infrared imager with 0.117 arcsec pixels and $2 \times 2$
arcmin$^2$ field of view \citep{hodapp03}.  We obtained a total of
22,635 s integration over 2 nights in the $H$ filter centered at 1.65
\micron\ (1.49 to 1.78 \micron).  The weather was photometric and the
seeing was good for both nights.  On the first night, the image
quality of the final image was 0.40 arcsec FWHM for 11,340 s of
exposure.  The second night was less good, and the overall image with
11,295 s of total exposure had 0.53 arcsec FWHM.  The images were
centered on the location of the $z{=}10$ candidate, rather than the
center of the cluster (as was the \citeauthor{pello04} data), to allow
better sky subtraction.

Individual 45 s exposures were reduced using the Gemini IRAF package.
Each frame was sky-subtracted and flattened using flat field
images produced by the Gemini calibration unit.  A unique sky image was
derived for each individual exposure by averaging the nearest 8 to 11
frames taken within 6 min, with stars and galaxies masked out.  Since the
$z{=}10$ candidate is located in a relatively low density region of the
cluster, residual background variations due to incomplete masking
of extended galaxies are not severe.  Individual sky-subtracted and
flat-fielded frames were then registered using integer shifts, averaged
without any weighting, and cosmic rays were rejected.  Data from the
two nights were reduced separately, and later combined to produce the
final image.  The image quality in the combined image is 0.47 arcsec FWHM.
Figure 1 shows the region around the candidate object.

Two UKIRT faint standard stars \citep{hawarden01} were observed and
reduced in the same way as the cluster data to determine the photometric
calibration.  The calibration determined for these two nights agreed to
within 0.01 mag with the extinction-corrected zero point over the two
month period prior to our observations.  We obtained a zero point at
H-band of 24.146 (compared to the NIRI mean of H=24.14 measured over
many months).  Our measured average sky brightnesses were H=14.15 and
14.27 mag per square arcsec for the two nights (typical for Mauna Kea).
The sky variations throughout both nights were a smooth function of
airmass, with the vast majority of data frames having values within 0.1
magnitudes of the above averages.  In addition, NIRI uses Hall effect
sensors to identify which filter is in the telescope beam. These identify
in a unique and absolute way a given filter making it impossible to have
an inappropriate filter in the beam.  The zero points and sky brightnesses
are in excellent agreement for NIRI in the H-band and are inconsistent
with any filter other than H.

For comparison, we retrieved the raw ISAAC/VLT data used by
\citeauthor{pello04}from the ESO data archive and re-reduced their
$H$-band data.  The resulting image is also shown in Figure 1 with the
same display parameters as the NIRI/Gemini image.  The final ISAAC
$H-$band image was made from the same $H-$band data as presented in
\citeauthor{pello04}as judged by comparing the total integration time
listed in their paper with the total integration time of data we
reduced (i.e., all the data from ESO programme ID 70.A-0355).

\subsection{Photometry}

To assess our ability to reliably detect objects with $H_{\rm
AB}\,{=}\,25.0$ \citep[the magnitude of the $z=10$ candidate measured
by][]{pello04}, we added twenty-eight artificial stars with $H_{\rm
AB}\,{=}\,25.0$ and 0.47 arcsec FWHM to the final Gemini image, three
of which are shown in Figure~2. The mean recovered magnitude for these
objects was $H_{\rm AB}\,{=}\,25.035$, with 1$\sigma$ uncertainties
on each measurement typically 0.18 magnitudes.  The aperture used for
photometry was a circle of diameter 3 times the seeing disk (a diameter of
1.4 arcsec). The good agreement between input and recovered magnitudes
indicate that no aperture correction is necessary.  The artificial
sources are similar in brightness to the four faint objects marked in
Figure~2 and suggest what the \citeauthor{pello04} object should look
like if it were present in our image. All of the objects were easily
detected in the NIRI image (see Figure~2).

Despite this, our Gemini $H$-band image shows no sign of an object at
the location of the $z{=}10$ galaxy candidate.  Given the measured sky
noise and the results of our artificial star analysis, a point source with
$H_{\rm AB}\,{=}\,25.0$ would have been detected with a $S/N\,{\sim}\,7-8$
in a photometric aperture three times the seeing FWHM (1.4 arcseconds
diameter). We randomly placed twenty-five 1.4 arcsec diameter apertures
on sky regions within the target area shown in Figure~1 in order to
determine detection limits for the candidate. The distribution of residual
flux in these apertures gave a $3-\sigma$ limit of $H_{\rm AB} >26.30$,
a result which agrees with the measured uncertainties on the magnitudes
of the artificial sources.

We also measured the distribution of the individual pixel counts about
the mode of the sky value for the entire frame, after excising pixels
containing flux from identified objects. This more conservative approach
takes into account systematic variations in sky level due to subtracting
the individual unique sky images, which are inevitably influenced by
small amounts of flux from objects not fully masked out when creating
the images. The resulting $3-\sigma$ limit of $H_{\rm AB} >26.03$
differs from the aperture determination by 0.27 magnitudes, indicating
that these systematic uncertainties are minor. Although, as noted above,
these systematic uncertainties are likely to be less of an issue in
the region of the candidate than for the frame as a whole, for the rest
of this paper we use the more conservative limit of $H_{\rm AB} >26.0$
for the flux at the position of the $z{=}10$ candidate.

Several objects detected in the region of our image shown in Figure~2 have
magnitudes comparable to or somewhat fainter than \citeauthor{pello04}'s
$H$-band magnitude for the candidate $z{=}10$ galaxy.  Four of these
objects lie within 15 arcsec of the $z{=}10$ candidate and provide
a robust indication of the depth of the image.  Object brightnesses
were determined using the IRAF apphot aperture photometry package.
All four have measured $S/N\,{\gtrsim}\,5$ within an aperture 1.4 arcsec
in diameter.  The positions of these four comparison objects are shown
in Figure~2, and their brightnesses are listed in Table~1.

\section{Discussion}
\label{discuss}

The lack of an $H-$band detection in our Gemini North NIRI data at a
level significantly fainter than the detection by \cite{pello04} is
puzzling. Our observations are deeper (by about one magnitude), and
have better spatial resolution and sampling. It is possible that the
object is time-variable or transient, in which case non-concurrent
multi-band photometry does not constrain the redshift, or has a large
proper motion ({\it i.e.}, is a solar system object: the ecliptic
latitude of the source is about 14 degrees) given the range of dates
over which the ESO data were obtained and the time elapsed between the
Gemini and VLT observations.

In any event, our non-detection in $H$ greatly weakens the argument
based on the large break between the optical, $J$, and $H$ bands which
supported the claim that the line detected in the spectroscopy is most
likely Ly$\alpha$ at $z=10.0$. Given the photometry in $J$ reported
by \citeauthor{pello04}, there is no formal continuum detection in
the optical or $J$. The photometry by \citeauthor{pello04}indicates
the object is only marginally detected in $H$ ($\sim$4$\sigma$;
$H_{AB}=25.00\pm0.25$) and $K$ (3$\sigma$; $K_{s,AB}=25.51\pm0.36$). With
our new $H-$band data ($H_{AB}>26.0$), the broad-band photometry can no
longer be said to constrain the redshift of this object at all.

This leaves the $\sim$4$\sigma$ detection of the emission line from
the spectroscopy. Recently \cite{Weatherley04} have thrown doubt on
the reality of this line. Taken together with our work, the reality
of any source at this position has to be strongly questioned. However,
whereas our work is based on an independent data set, \cite{Weatherley04}
reanalyse the \citeauthor{pello04} data. It is possible that small
differences in reduction may lead to contentious results, especially
given the faintness and low signal-to-noise of the detection claimed by
\cite{pello04}.  Consequently, the following discussion assumes that the
line is real. But even if it is not, the discussion is directly relevant
to all searches for high redshift line emitters.

It is still possible that this line is Ly$\alpha$ at $z{=}10$, but with
only this level of detection and no corroboration from photometry,
the case is weakened considerably and perhaps this is not the most
likely interpretation.  Other lines such as [OII], [OIII], H$\alpha$
at redshifts between $0.77<z<2.75$ were detectable given the wavelength
range used in the spectroscopy.  The $H-K_s$ color can no longer be
used as evidence for an intrinsically blue and hence extremely young
high redshift galaxy, given our sensitive upper limit.

Luminosity distributions for high redshift objects in \cite{SLBGs4}
indicate that there should be several tens of low continuum luminosity
galaxies per square arcminute between $0.77<z<2.7$ with continuum
fluxes below the optical flux limits quoted in \cite{pello04}.  Indeed,
\cite{richard03} discovered a faint $z=1.7$ galaxy using the same imaging
and spectroscopic data as used by \citeauthor{pello04}  This object
was identified in continuum longward of the $I$ band from imaging and
in three emission lines from spectroscopy. A comparable galaxy several
times fainter in both continuum and lines would have been undetected
in the photometry and would have only been detected in a single narrow
line. Thus it remains possible that the \citeauthor{pello04}object is
a galaxy of this type, almost regardless of which of the above emission
lines is the true identification.

HII dwarf galaxies in the local universe and at moderate redshifts
show a correlation between H$\beta$ luminosity and emission line
widths \citep{melnick00}.  \citeauthor{pello04}measure a line flux of
$\sim$4$\times$10$^{-18}$ ergs s$^{-1}$ cm$^{-2}$ and an upper limit to
the line width of about 60 km s$^{-1}$. If we assume the galaxy to be at
redshift appropriate for the most likely alternative line identifications
such as [OII]$\lambda$3726, [OII]$\lambda$3729, [OIII]$\lambda$5007
or H$\alpha$, a reasonable range of ratio of H$\beta$ to these other
lines for dwarf galaxies, and a small amount of magnification due to the
intervening cluster, we find that the emission line plausibly lies along
the correlation of \cite{melnick00} for line width versus H$\beta$ line
luminosity. As an example, suppose that the actual line identification
is [OIII]$\lambda$5007 at a redshift of 1.67, leading to a luminosity
of 40.9 ergs s$^{-1}$ in the log assuming reasonable cosmological
parameters (H$_0$=70 km s$^{-1}$ Mpc$^{-1}$, $\Omega_{matter}$=0.3
and $\Omega_{\Lambda}$=0.7).  For HII/dwarf galaxies, the ratio of
[OIII]$\lambda$5007 to H$\beta$ ranges from about 2 to up to 10,
consequently the H$\beta$ luminosity would be between 39.9 to 40.6
ergs s$^{-1}$ in the log. For the published line width, this range in
luminosity falls on the correlation between line luminosity and width
given by \citeauthor{melnick00} for HII/dwarf emission line galaxies,
showing that identifying this line as an optical emission line from a
moderate redshift HII/dwarf galaxy is plausible.

\acknowledgments

We wish to thank the pair of anonymous referees for their insightful and
encouraging comments, and Roser Pell\'{o}, Daniel Schaerer, and Dan Stern
for their helpful comments.  These results are based on observations
obtained at the Gemini Observatory, which is operated by the Association
of Universities for Research in Astronomy, Inc., under a cooperative
agreement with the NSF on behalf of the Gemini partnership: the National
Science Foundation (United States), the Particle Physics and Astronomy
Research Council (United Kingdom), the National Research Council (Canada),
CONICYT (Chile), the Australian Research Council (Australia), CNPq
(Brazil) and CONICET (Argentina). These Gemini observations were supported
by Director's Discretionary Time.  The data retrieved from the ESO data
archive were from observations made with the Paranal Observatory under
programme ID 70.A-0355(C). LD acknowledges receipt of a PPARC studentship.

\begin{deluxetable}{cc}
\tablewidth{150pt}
\tablecolumns{2}
\tablecaption{Aperture Photometry of Objects in the Abell 1835 Field}
\tablehead{
\colhead{Object} &
\colhead{$H_{\rm AB}$}
\\
\colhead{ID\tablenotemark{a}}&
\colhead{(AB mag)}
}
\startdata
Cand  & $>26.0$\tablenotemark{b} \\
1     & $25.26 \pm 0.21$ \\
2     & $25.23 \pm 0.21$ \\
3     & $24.70 \pm 0.14$ \\
4     & $25.17 \pm 0.20$ \\
Art 1& $25.04 \pm 0.18$  \\
Art 2& $25.00 \pm 0.17$  \\
Art 3& $25.06 \pm 0.18$  \\

\enddata \tablenotetext{a}{Artificial objects all have input $H_{\rm
AB}\,{=}\,25.0$.}  \tablenotetext{b}{3-$\sigma$ upper limit}
\tablecomments{All magnitudes and limits measured in 1.4 arcsec
diameter apertures. Uncertainties were determined from measurements of
rms pixel-to-pixel variation in the sky level directly around each
aperture and across the frame as a whole, the quoted value being the
larger of these two in each case.}

\end{deluxetable}

\clearpage

\begin{figure}
\includegraphics[scale=0.9]{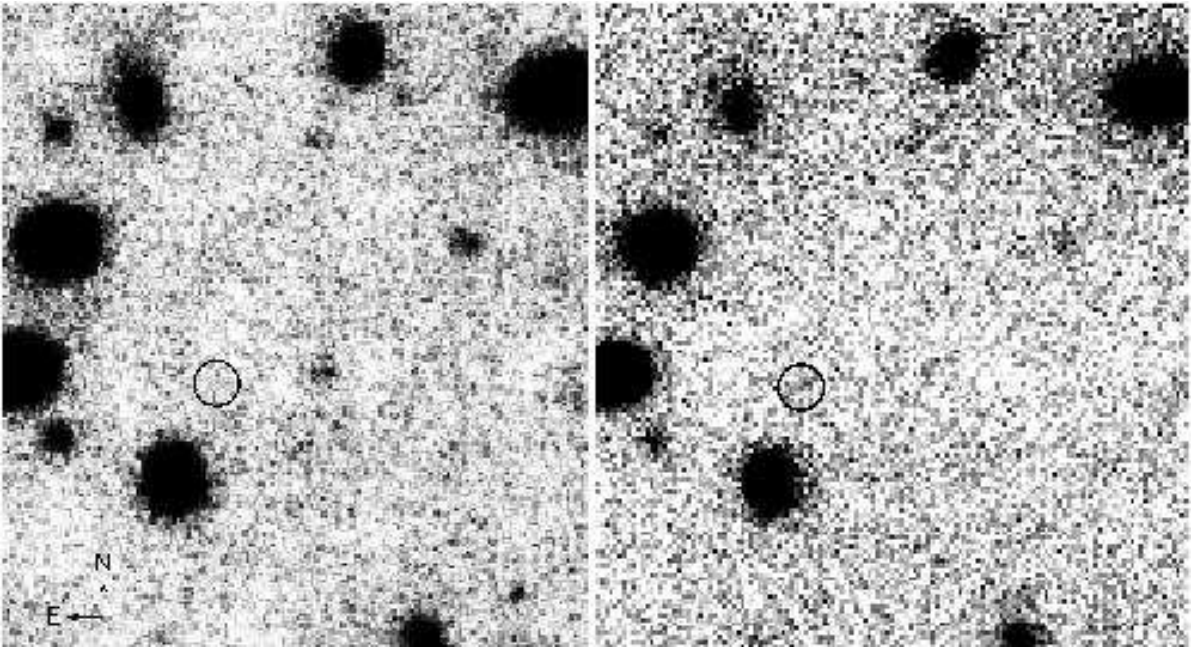}
\caption{Deep $H$-band NIRI image of the region around the $z{=}10$
candidate object (left) and our reconstruction of the ISAAC/VLT $H$-band
image (right).  Both images are 18 by 22 arcsec, the NIRI image has
0.47 arcsec FWHM image quality, and both images are displayed with
the same linear range and stretch with respect to their noise levels.
The circles indicate the location of the candidate object, and are 1.4
arcsec in diameter.  A comparison of the aperture (1.4 and 3.0 arcsec in
diameter) magnitudes of 15 relatively bright (H$_{AB}$$\approx$19-23)
objects near the position of the candidate object (estimated using
the average zero-point from the ESO web pages for ISAAC for the month
around the time of the observations) suggests that the photometry between
the ISAAC and NIRI images are consistent with an average difference of
$<$H$_{AB,ISAAC}$-H$_{AB,NIRI}$$>$=$-$0.09$\pm$0.10 magnitudes, with no
dependence on the $J-H$ color of an object.}
\end{figure}

\begin{figure}
\epsscale{0.7}
\plotone{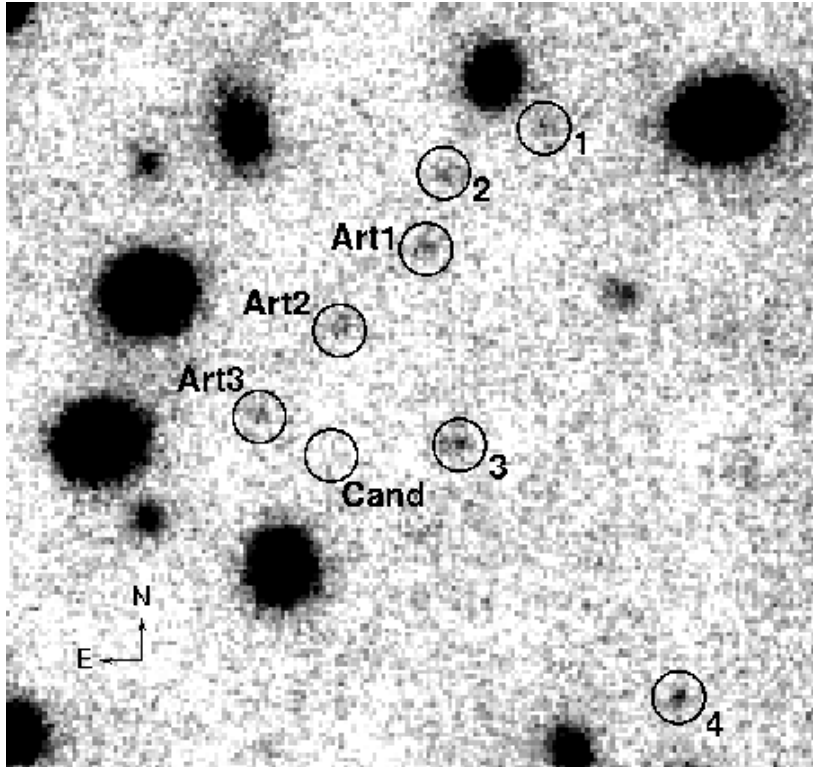}
\caption{The same Gemini data as in Figure 1, this time with
positions of four other sources near the candidate object with $H_{\rm
AB}\,{\gtrsim}\,25$. These are barely visible in the VLT data shown
in Figure 1.  In addition, the objects labelled ``Art'' are artificial
sources with $H_{\rm AB}\,{=}\,25.0$. The three artificial sources are
an illustrative subset of 28 artificial point sources used to assess the
detectability of objects with this brightness in our image. The photometry
for sources 1-4, and these three artificial sources is given in Table 1.}
\end{figure}

\end{document}